# Motion Detection Notification System by Short Messaging Service Using Network Camera and Global System for Mobile Modem

Mohd Norzali Haji Mohd, Mohd Helmy Bin Abd Wahab and Siti Khairulnisa Ariffin

**Abstract**— As the technology rapidly grows, the trend is clear that the use of mobile devices is gain an attention nowadays, thus designing a system by integrating it with notification feature is becoming an important aspect especially in tracking and monitoring system. Conventional security surveillance systems require the constant attention from the user, to monitor the location concurrently. In order to reduce the cost of computing power and advance technology of mobile phone in widespread acceptance of the Internet as a viable communication medium, this paper is aimed to design a low cost web-based system as a platform to view the image captured. When the network camera detects any movement from the intruders, it automatically captures the image and sends it to the database of the web-based directly by the network through File Transfer Protocol (FTP). The camera is attached through an Ethernet connection and power source. Therefore, the camera can be viewed from either standard Web browser or cell phone. Nowadays, when a security camera is installed, user is notified as long as the camera is switched on since any slight movement requires the attention of the supervisor. The utility of the system has proven theoretically. This system will also notify the user by sending a notification through Short Messages Services (SMS).

## 1 INTRODUCTION

MOTION Detection Notification System using SMS and Network Camera is full-featured with surveillance system that connects to the Ethernet network to provide remote high-quality audio and video. It comes with auto-motion caption, capture in extreme low light environment and others special specifications. It can detect motion, act or process of changing position or place. The basic idea is simple: to detect motion in a scene where there would normally be no motion. The simplest way of performing this is by differentiating successive images, or comparing successive images with a template of the scene [7]. Then, it produces a signal such as sound or feedback to the user once surveillance occurs. The system is also known as alarm security system, surveillance system, burglar system or circuit close television (CCTV). Generally, there are five types of security camera namely indoor camera, outdoor camera, pan-tilt-zoom camera, night vision camera and hidden camera.

The use of this system is for the protection of property and people which becoming more widespread due to the cost advantage over typical perimeter and area detection devices. Generally, this security system can detect a motion when the motion grade has reached a certain level set in the system

This work is supported by UTHM short grant scheme (Vote number: 0618).

- *Engr.Mohd Norzali Mohd is a Lecturer with the Universiti Tun Hussein Onn Malaysia*
- *Mohd Helmy Bin Abd Wahab is a Lecturer with the Universiti Tun Hussein Onn Malaysia*
- *Siti Khairulnisa is a graduate student at UTHM*

### 1.1 Motion based detection

Motion Detection approaches used in current market. Whereas, motion detection is one of the most important and interesting issues in the area of image processing. Motion detection includes methods by which motion has been identified. Motion detection sensor are available which can easily detect sound and produce an alarm or switch on an image recording device.

Motion based detection relies entirely on moving aspects of object to detect its movement as shown in Figure 1. One of the methods of motion detection based on the second approach is a temporal derivative technique [10].

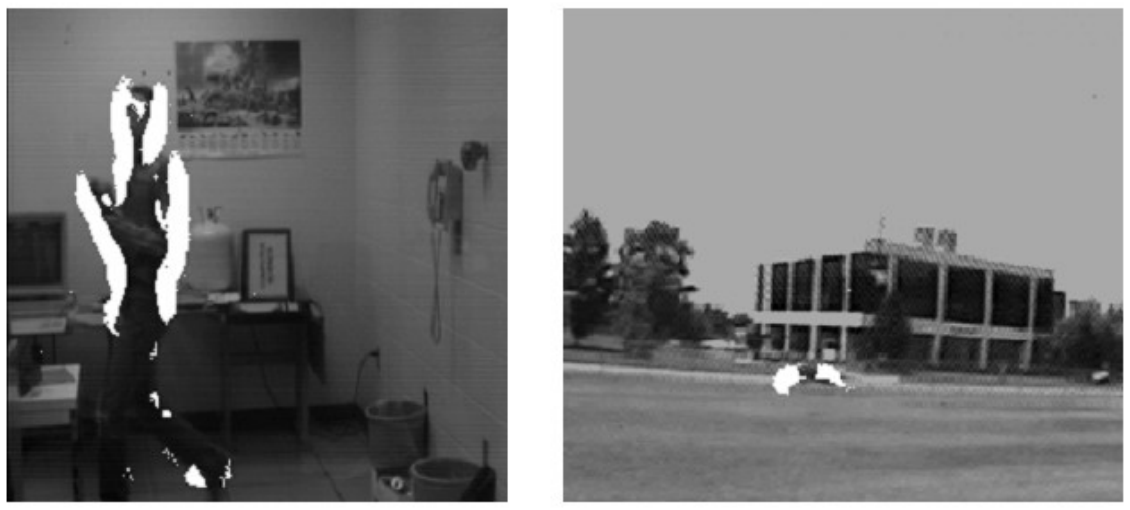

Figure 1 Example shot of motion based detection

### 1.2 Pyroelectric infrared sensor

The camera features a built-in pyroelectric infrared sensor, which uses infrared rays to detect temperature differences within its range that are emitted naturally by people.

The pyroelectric sensor is made of a crystalline material that generates an electric surface charge when exposed to heat in the form of infrared radiation. When the amount of radiation striking the crystal changes, the amount of charge also changes and can be measured with a sensitive FET device built-in in the sensor.

The PIR sensor has two sensing elements connected in a voltage bucking configuration. This arrangement cancels signals caused by vibration, temperature changes and sunlight. A body passing in front of the sensor will activate first one and then the other element whereas other sources will affect both elements simultaneously and be cancelled. In Figure 2, the radiation source must pass across the sensor in a horizontal direction when sensor pins 1 and 2 are on a horizontal plane so that the elements are sequentially exposed to the IR source.

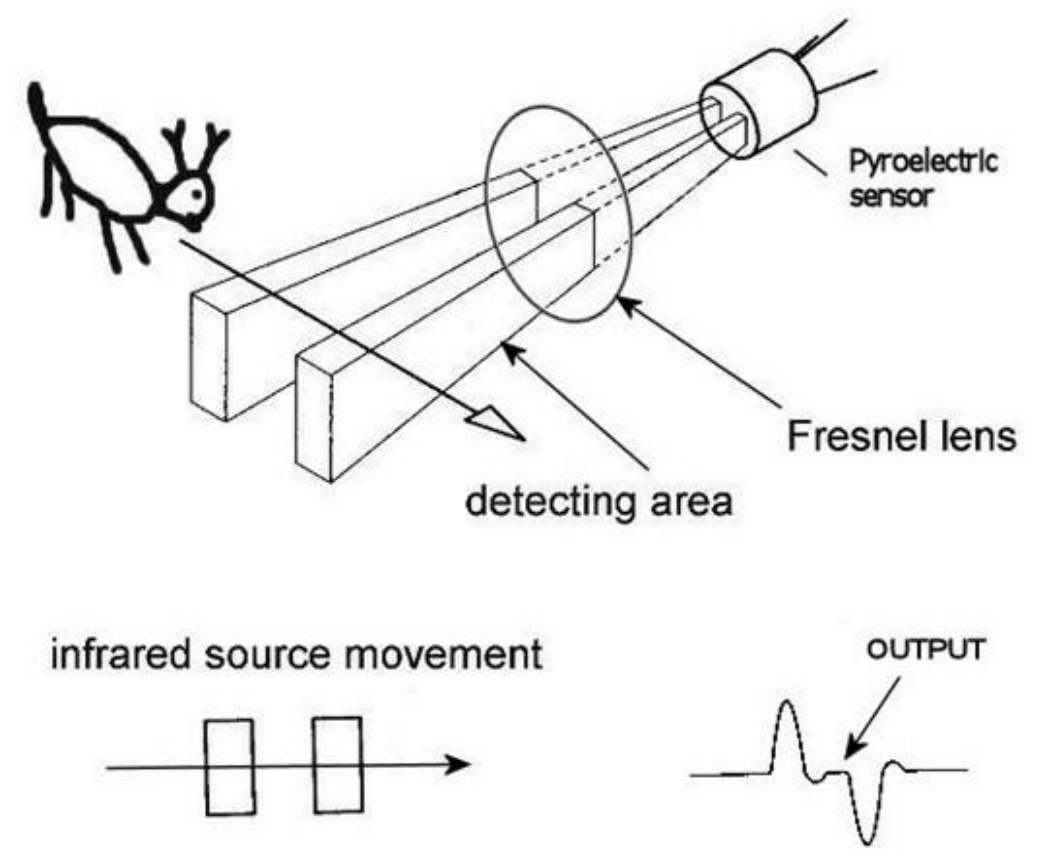

Figure 2 Pyroelectric Function

## 2 METHODOLOGY

Generally, the system was divided into two phase; hardware design and software development.

### 2.1 Hardware Design

In this system, a network camera transmits image data and control signals over a Fast Ethernet link. When image is captured, it transfers to the database. GSM modem is integrated to the system to send SMS when a picture is detected in database. The flow is considered as in Figure 4.

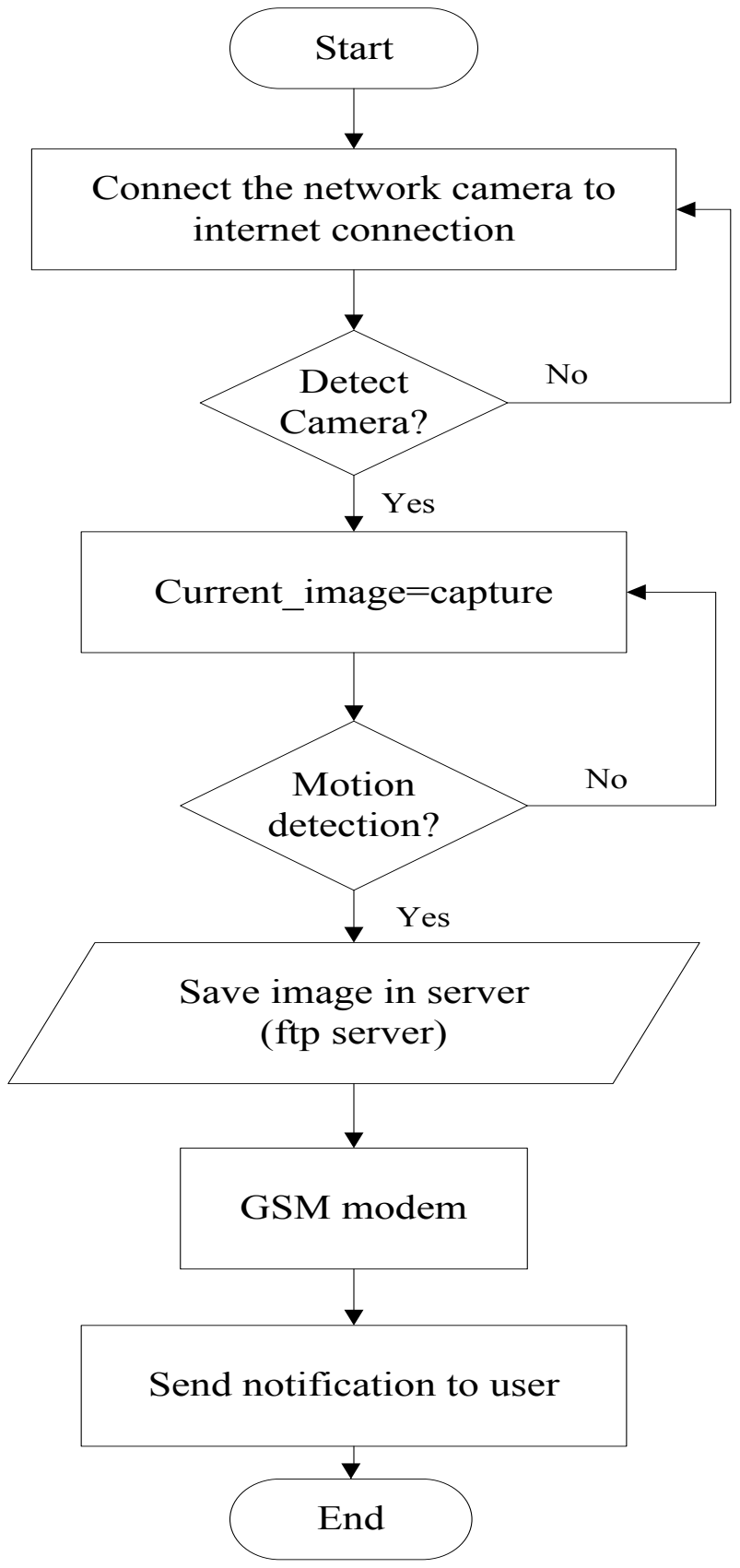

Figure 4 Hardware Configuration

For motion detection using network camera, a threshold determines at what point the motion detection feature is triggered. A lower threshold means less motion is needed to trigger the motion detection feature. A higher threshold means more motion is needed to trigger the motion detection feature. As illustrated in Table 1 and 2, threshold is indicated in the [Preview] area by the light green area.

While for sensitivity, it determines how easily the camera detects motion. Lower sensitivity means the camera is less likely to detect motion. Higher sensitivity means the camera is more likely to detect motion. Motion is indicated in the [Preview] area in dark green when it has not exceeded the detection threshold, and is indicated in dark red when it has exceeded the detection threshold.

TABLE 1
LOW TRESHOLD SETTING

| Parameter | Display | Meaning |
|---|---|---|
| Threshold | 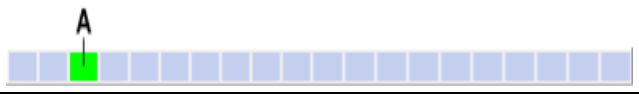 | Low threshold (A) |
| Sensitivity | 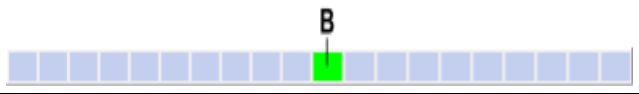 | Moderate sensitivity (B) |
| Preview | 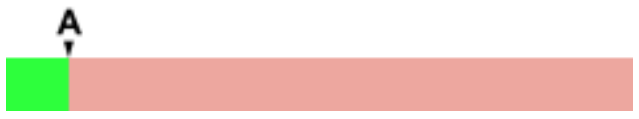 | Low threshold (A), no motion is detected |
| Preview | 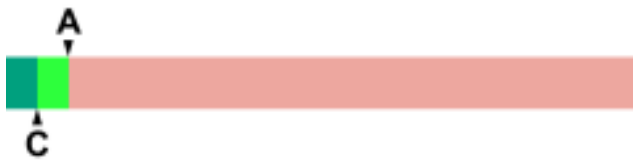 | Low threshold (A), low motion (C), motion detection is not triggered |
| Preview | 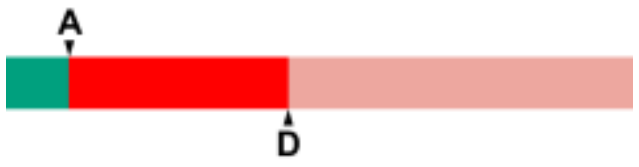 | Low threshold (A), moderate (D), motion detection is triggered |

TABLE 2
HIGH THRESHOLD SETTING

| Parameter | Display | Meaning |
|---|---|---|
| Threshold | 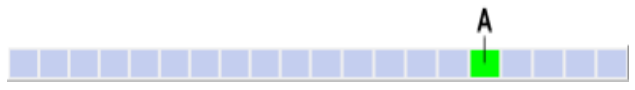 | High threshold (A) |
| Sensitivity | 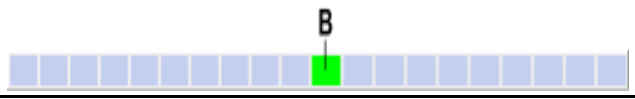 | Moderate sensitivity (B) |
| Preview | 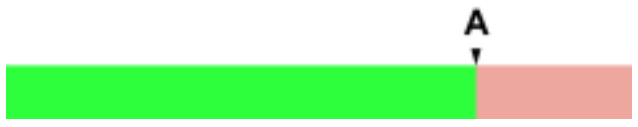 | High threshold (A), no motion is detected |
| Preview | 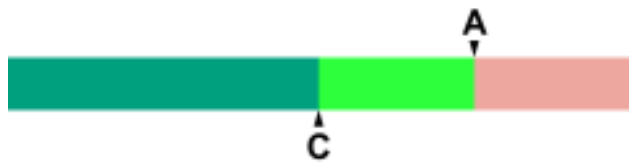 | High threshold (A), moderate motion (C), motion detection is not triggered |
| Preview | 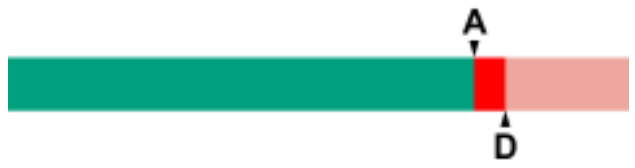 | High threshold (A), high motion (D), motion detection is triggered |

Figure 5 shows how the GSM modem works with the Web-based system to send an SMS to the user.

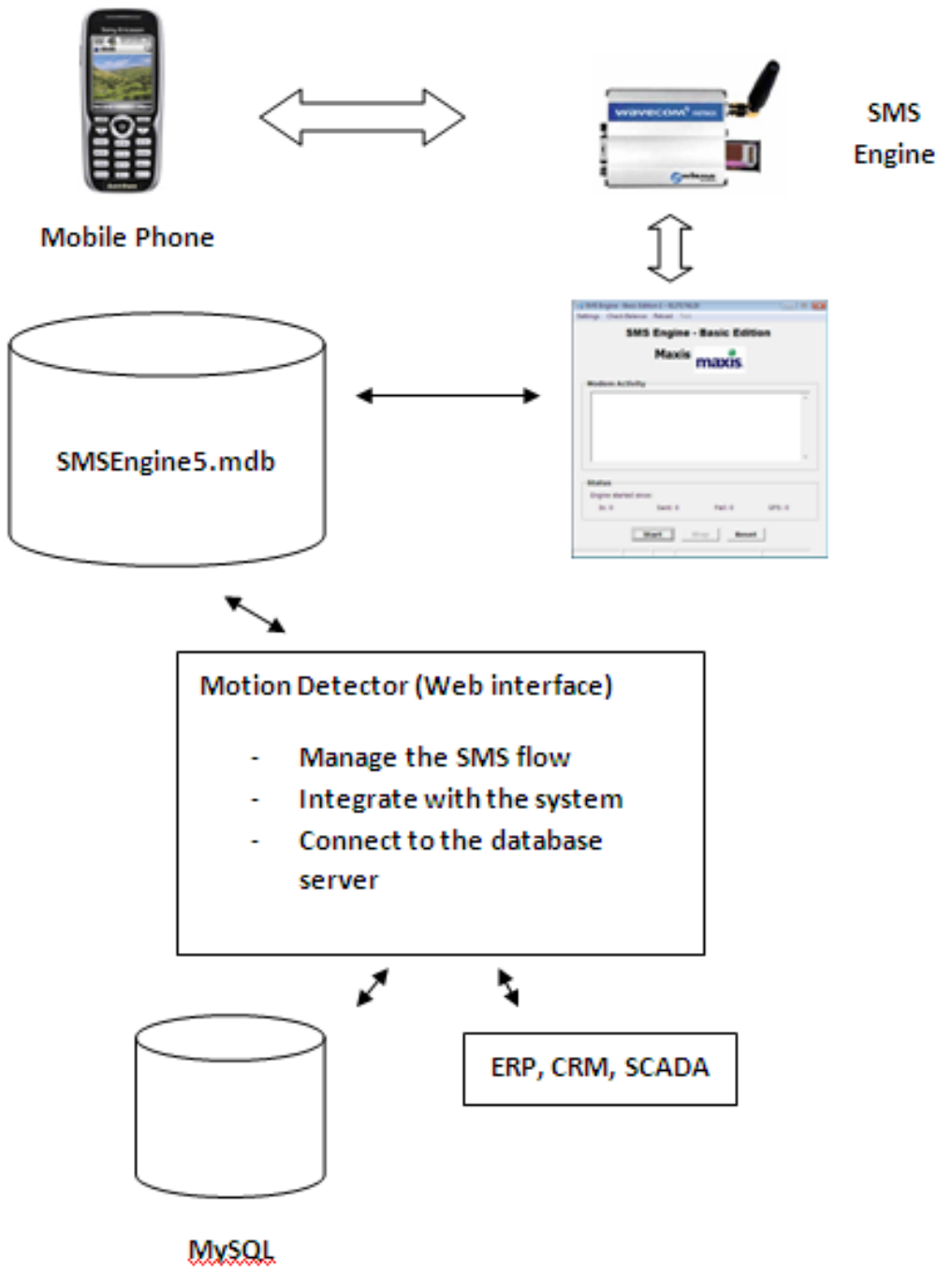

Figure 5 Interactions of Web Interface and SMS Engine

# 3 RESULT AND ANALYSIS

The prototype is illustrated in Figure 6 which consists of the integration between hardware and software. The details on how to install and test the hardware, and the interfacing programming by using PHP script is explained. For the hardware part, it consist the installation of Panasonic Network Camera and Fastrack GSM Modem to the computer via LAN cable, RS-232 serial cable and RS-232 USB serial port. While on the other hand, the software is developed by using PHP scripting language to create the web-based and installed using WAMPSERVER. Then the SMS is generated by PHP scripting using SMS Engine interface.

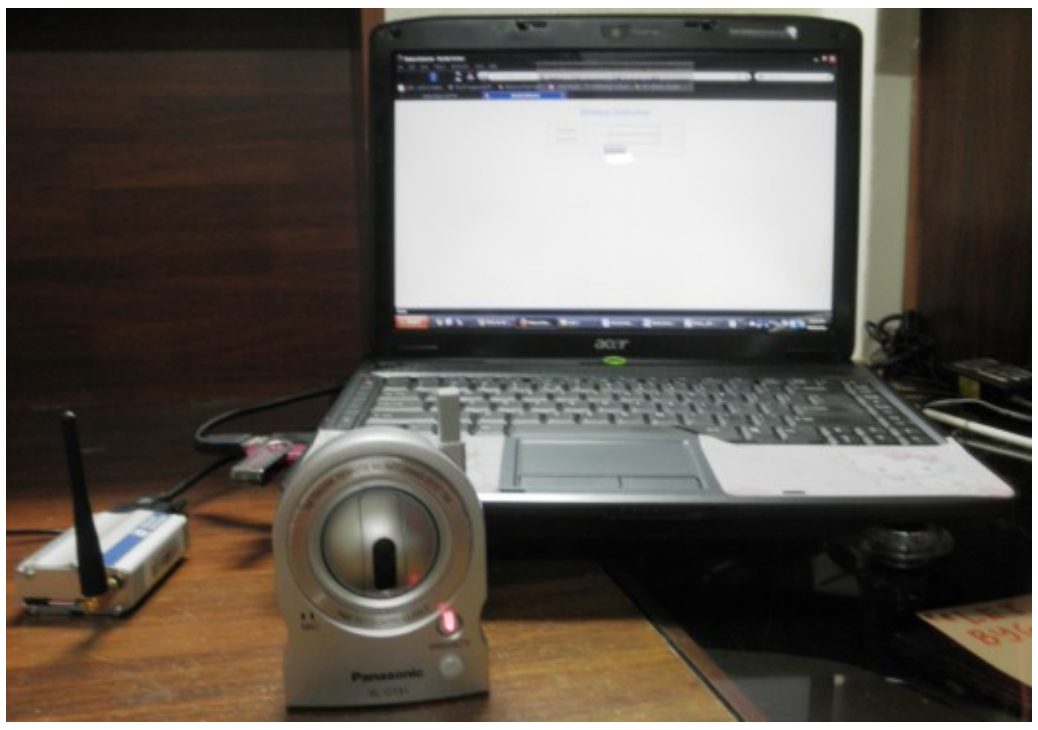

Figure 6 The Complete System

## A. Hardware configuration

SMS Engine

After the SMS Engine is finally set up, the log of the GSM modem can be viewed from the modem activity as in Figure 7 and figure 8.

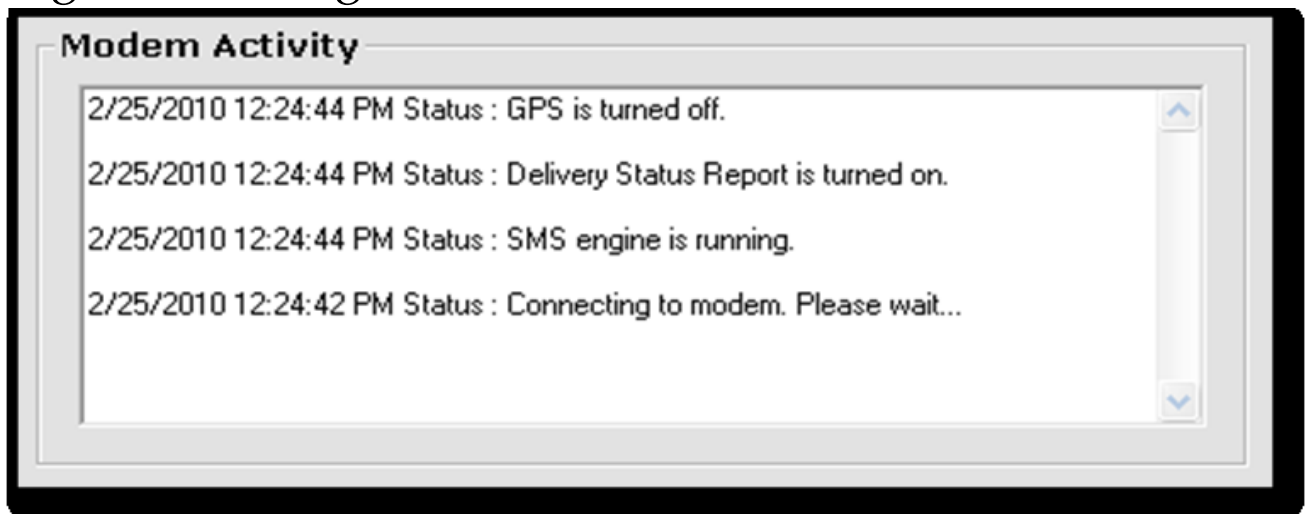

Figure 7 Log View of Modem Activity when SMS Engine is Start

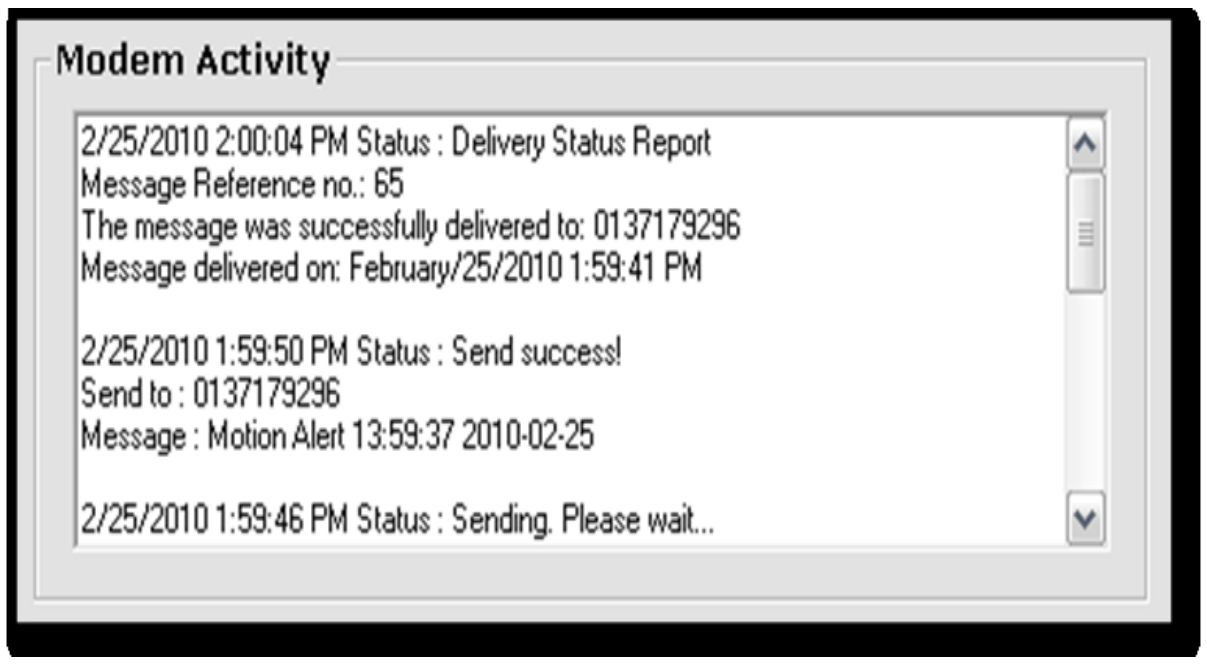

Figure 8 Log View of Modem Activity when SMS Engine is Successfully Send an SMS

Network Camera

In Figure 9, resolution for image Capture/Transfer is set so that the image can be viewed through web interface via mobile phone.

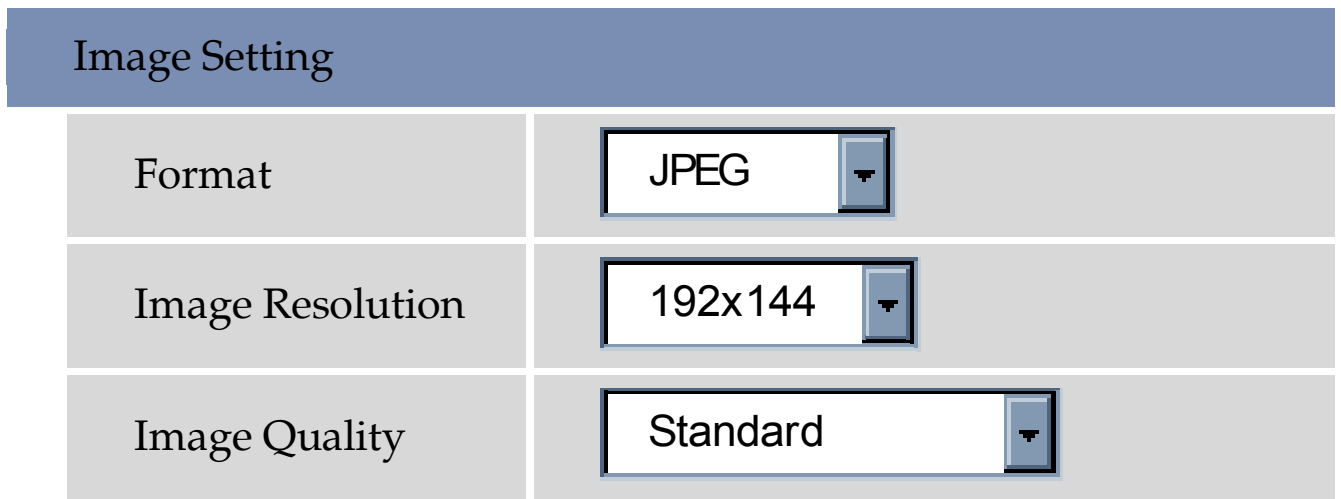

Figure 9 Image Setting

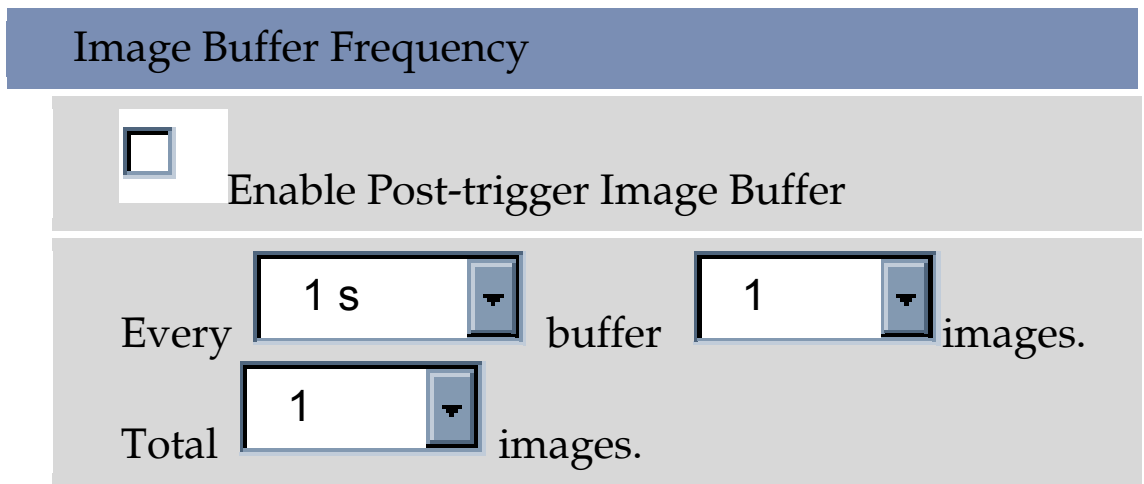

The Post trigger in Figure 10 can be set on the buffered image. Frequency and total number of images to Capture and Transfer also can be setup. Total number of images may vary depending on memory size, resolution, quality or photographic subject.

Figure 10: Image Buffer Frequency

Sensor deactivation time in Figure 11 is set to give a time interval for the network camera to detect and triggered the next image.

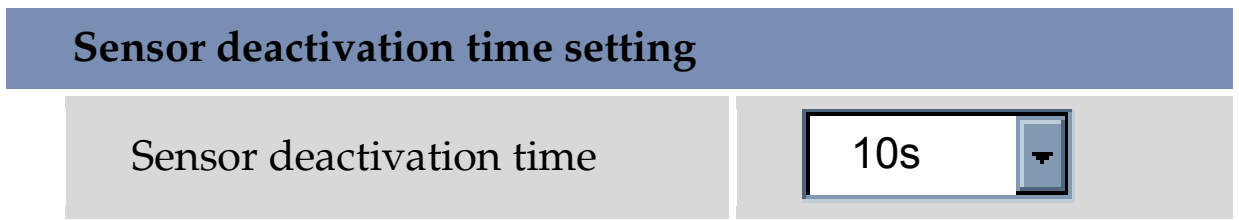

Figure 11 Sensor Deactivation Time Setting

In Figure 12, FTP is chosen as a transfer method. FTP server, port number, login ID, password in Figure 13 is set using the IP address that produce by WinFTP Server. The file name for the image is tested because already defined by PHP script, during the test image is transfer to the folder, the alert process can be executed.

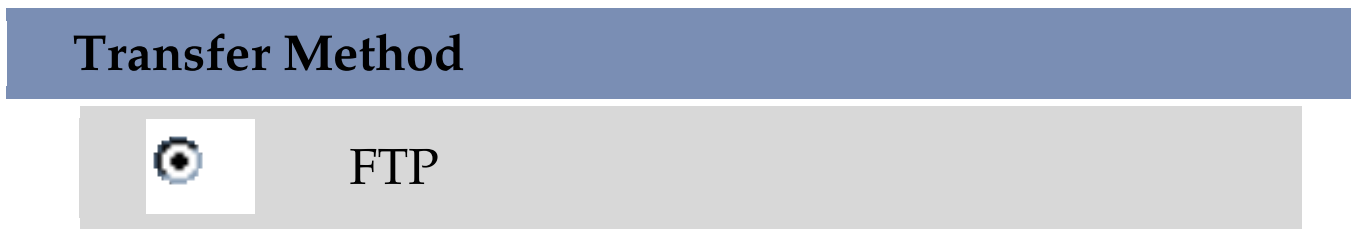

Figure 12: Transfer Method

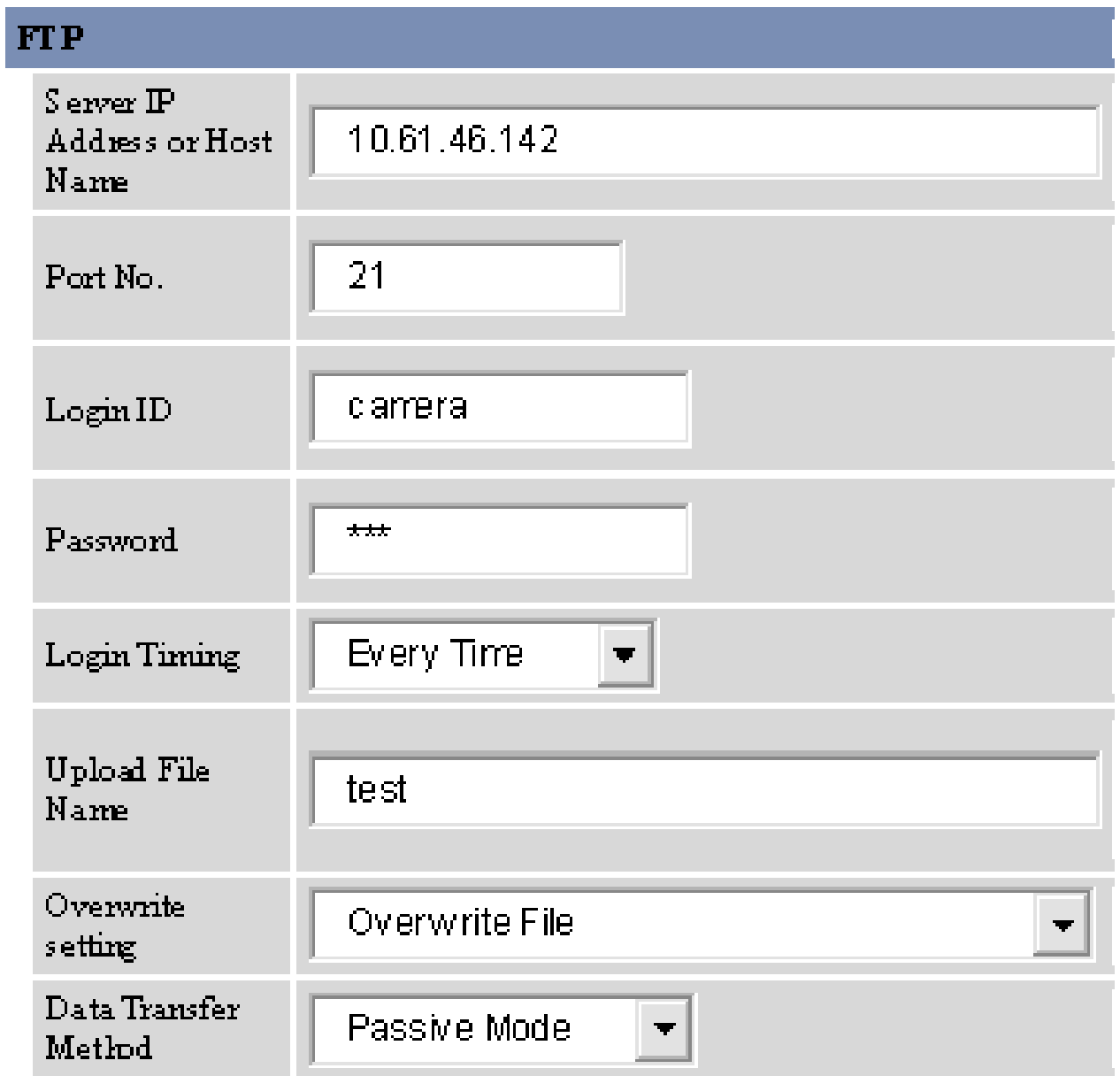

Figure 13: Transfer Image by FTP

## B. System Development

WampServer is required to host the web scripting which all the scripting code is place in WWW directory.

**Database using MySQL**

PhpMyAdmin is packec with wampserver to manage MySQL data. The database contains user account, user contact and user photo that is used for the web interface. Figure 14 demonstrates the details information of the configuration.

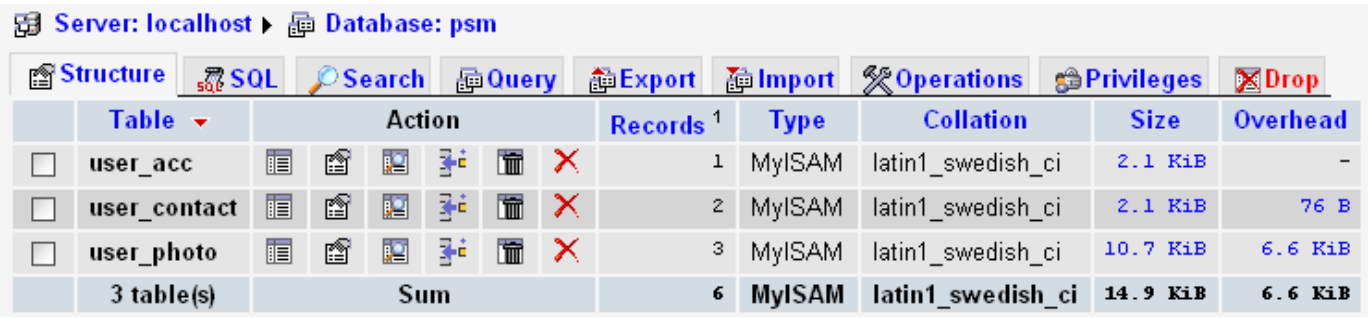

| | Table | Action | Records | Type | Collation | Size | Overhead |
|---|---|---|---|---|---|---|---|
| ☐ | user_acc | | 1 | MyISAM | latin1_swedish_ci | 2.1 KiB | - |
| ☐ | user_contact | | 2 | MyISAM | latin1_swedish_ci | 2.1 KiB | 76 B |
| ☐ | user_photo | | 3 | MyISAM | latin1_swedish_ci | 10.7 KiB | 6.6 KiB |
| | 3 table(s) | Sum | 6 | MyISAM | latin1_swedish_ci | 14.9 KiB | 6.6 KiB |

Figure 14 Database for the system

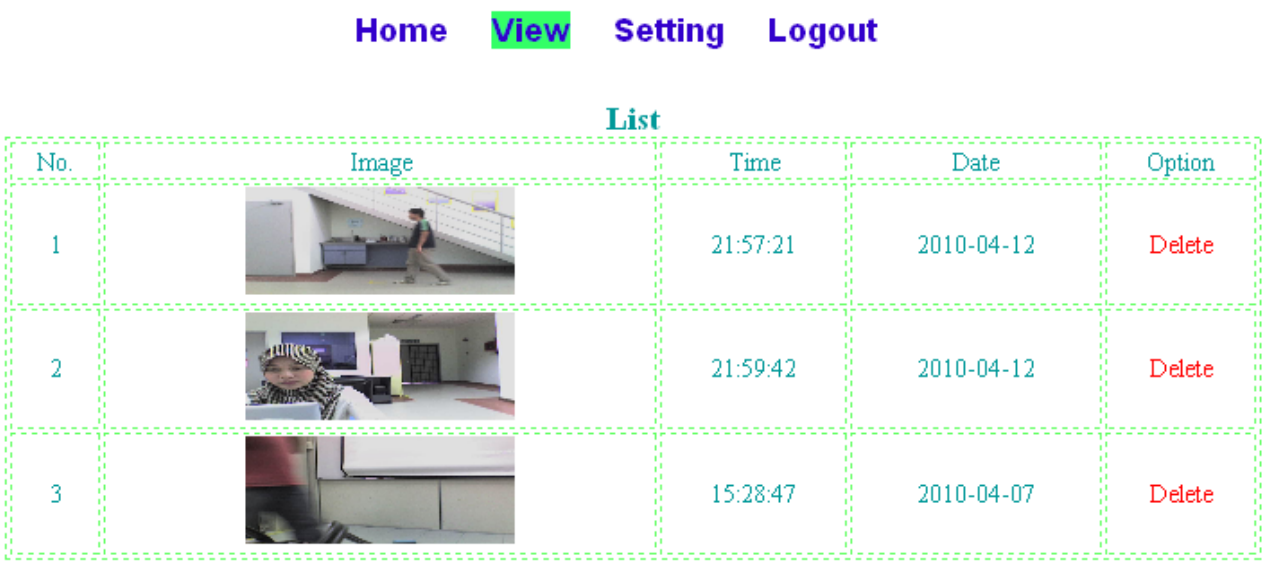

Figure 17 View Window

**Web Interface**

A graphical web page has been design using PHP script, to provide an interface for the user to access motion detection security system from any location in the world that has an Internet connection.

The web interface consists of four main screens: login screen (Figure 15), home screen (Figure 16), achieve screen (Figure 17) and setting screen (Figure 18). The login screen is designed for security purposes. The user logs in the web-based by supply the username and password. After authentication, the user is directed to the home screen where the latest picture captured by the network camera when motion is detected is display. Otherwise an error message is appeared.

To explore others images, thumbnail view screen is provided.

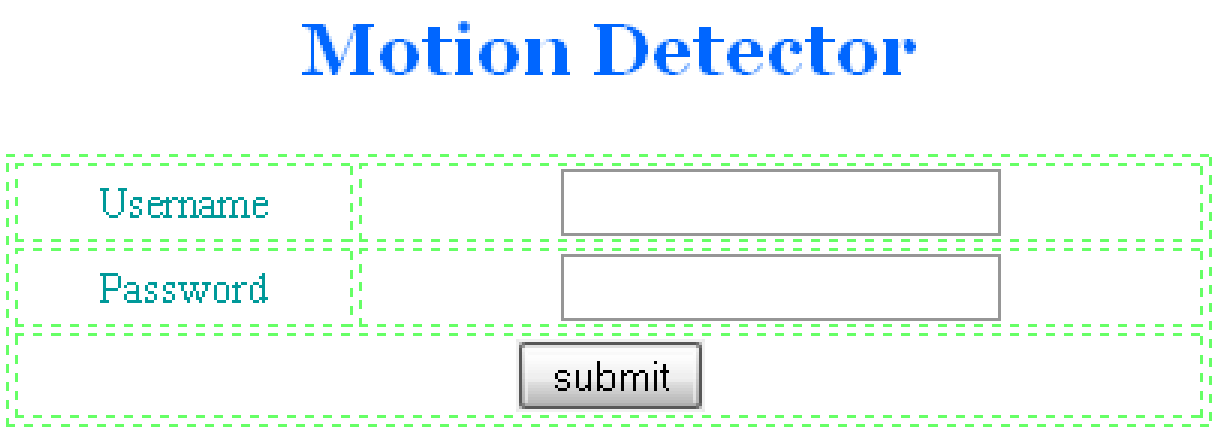

Figure 15 Login's window for the web interface

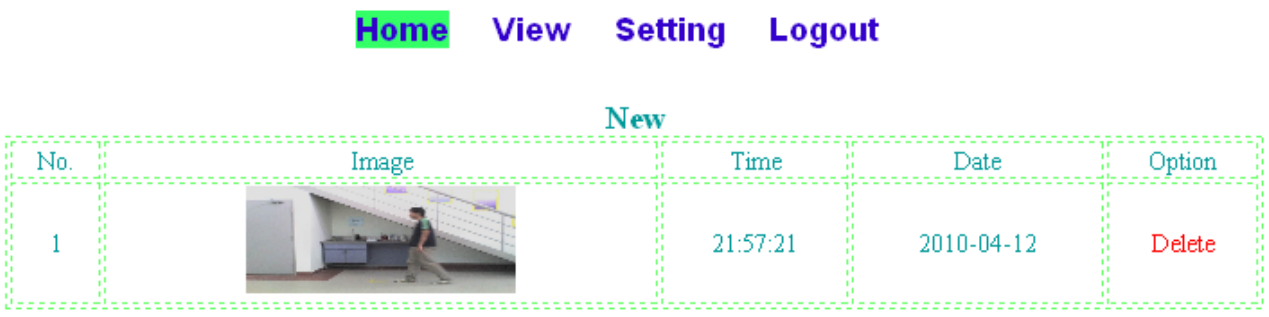

Figure 16 Home Window

Figure 17 shows View's window function as archive that contains the collection of pictures captures by network camera. User can use this window function to review back the old photos.

As depicts Figure 18, user can change the password anytime when login into the web interface. Furthermore, user also can add any mobile phone number to receive no

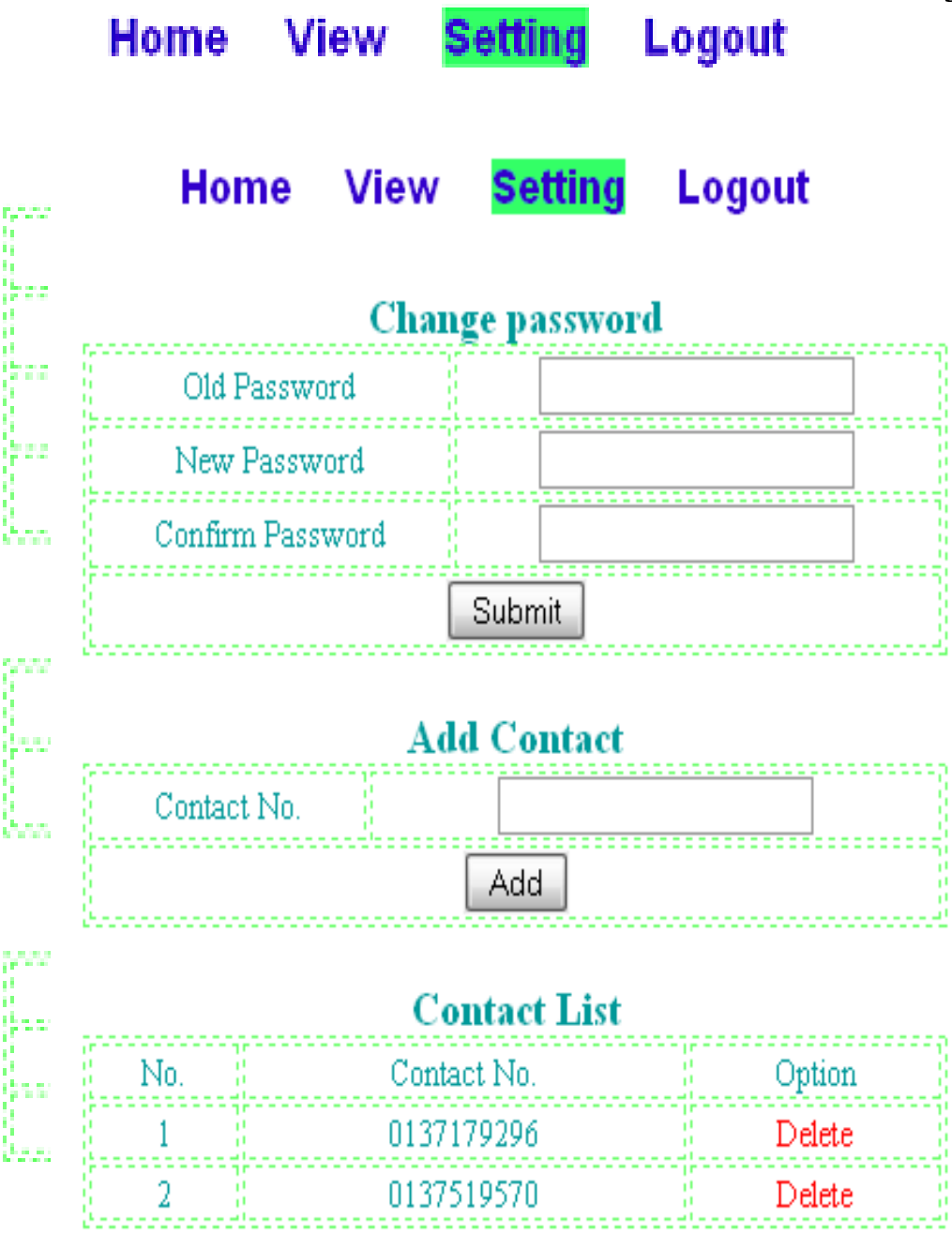

Figure 18 Setting Window

**FTP Server**

A WinFTP server software is used for file transfer. The user can created a FTP account by inserting username and password. Then a home directory is being selected for this user at C:\wamp\www\photo\source. When the WinFTP is put the server "Online", user can get the IP address for the ftp server.

In figure 19, Event Manager is used to execute alert_process to send an SMS (Figure 20). Execute features is selected to launch the coding in Figure 20 and to specify applications, the command %f: filename is used. So that during transferring the picture, the SMS Engine is execute to send SMS.

In Figure 21 shows that an SMS is successfully sent to the mobile number.

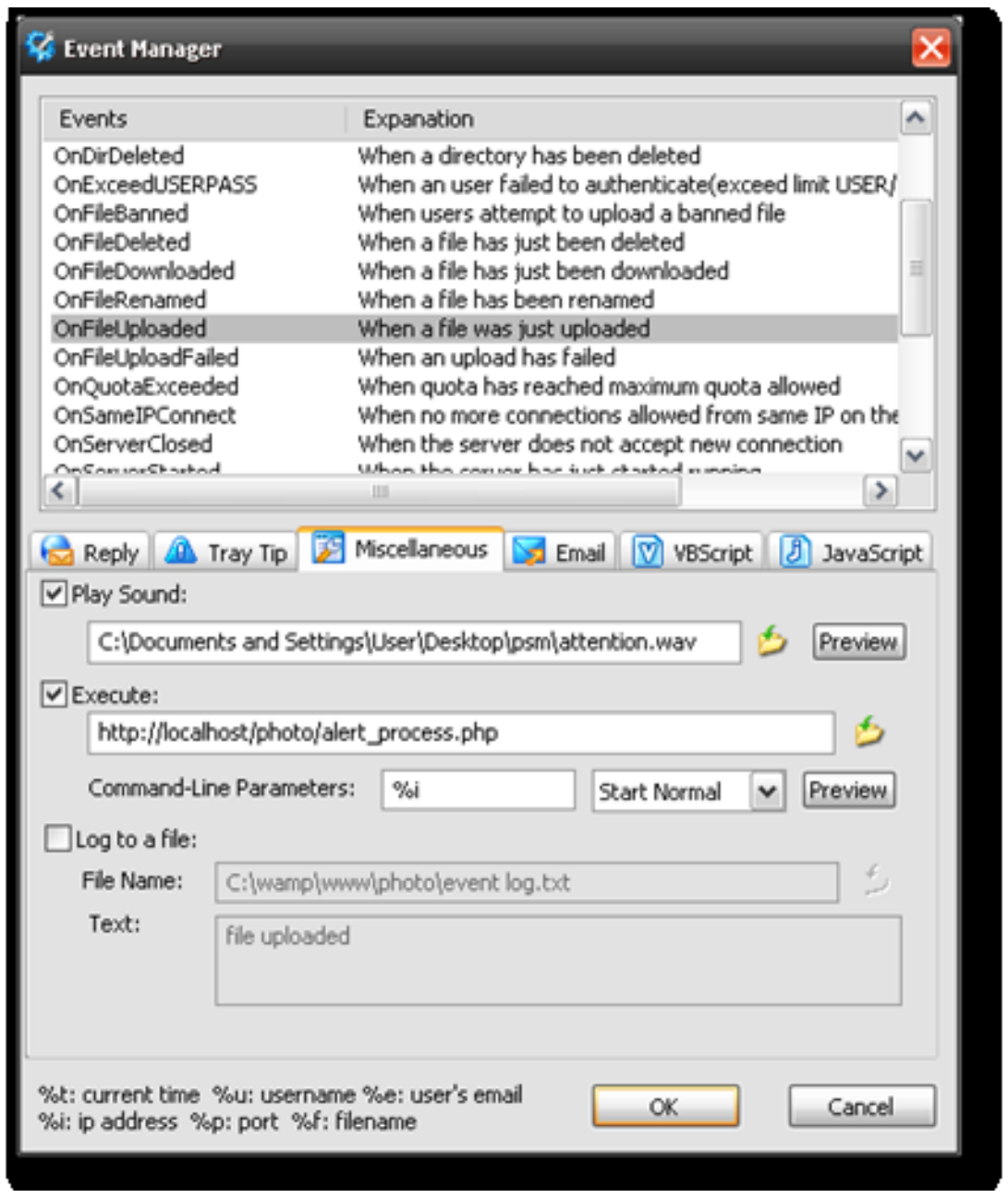

Figure 19 Event Manager Window

```
<?php
date_default_timezone_set('Asia/Kuala_Lumpur');
$today = date("Y-m-d");
$time = date("H-i-s");
$sql_time = date("H:i:s");
$file_name = $today . "_" . $time;
$file = 'source/test.jpg';
$newfile = 'image/' . $file_name . ".jpg";
if (!copy($file, $newfile))
{
echo "failed to copy $file...\n";
}
include('connection.php');
$sql = "INSERT INTO user_photo (photo_name,
photo_time, photo_date) VALUES('$newfile','
$sql_time', '$today')";
$query = mysql_query($sql) or die(mysql_error());
$gTime = $sql_time;
$gDate = $today;
$hp = "";
$sql = "SELECT * FROM user_contact;";
$query = mysql_query($sql) or die(mysql_error());
while($list = mysql_fetch_array($query))
{
$hp = $hp . $list['contact_no'] . "~";
}
$url = "SendSMS.php?contact=" . $hp . "&time=" .
$gTime . "&date=" . $gDate;
mysql_close($connect);
header('Location: ' . $url);
?>
```

Figure 20: PHP Scripting to Run Alert Process

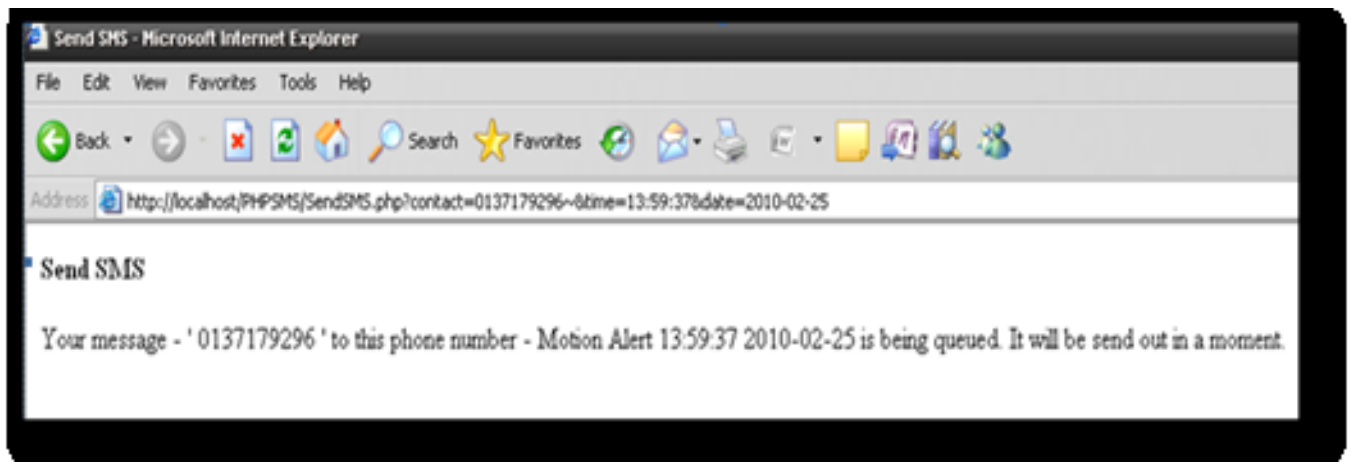

Figure 21 After SMS is Sent Successfully

Figure 22 below shows the SMS received by the user. The SMS contains time and date of the picture captured by the network camera.

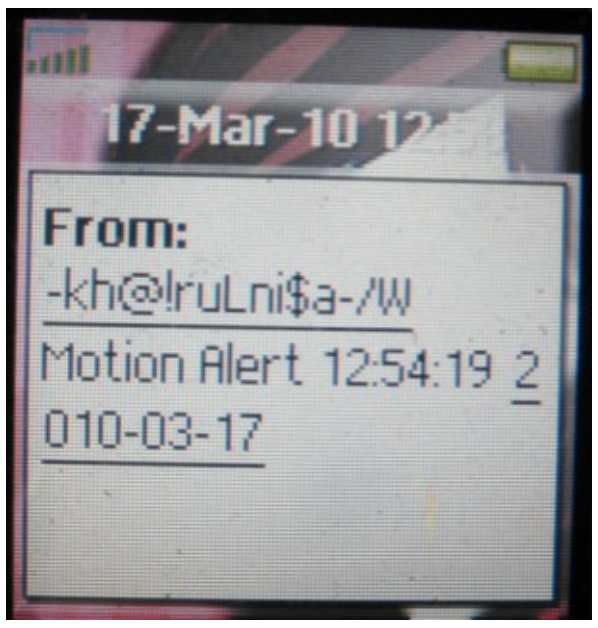

Figure 22 The SMS Received by the User

## A. Data Analysis

**IP Address**

From the FTP Server IP Address: 10.61.46.142, an IPv4 address has 32 bits long. The address space of IPv4 is $2^{32}$ or 4294967296.

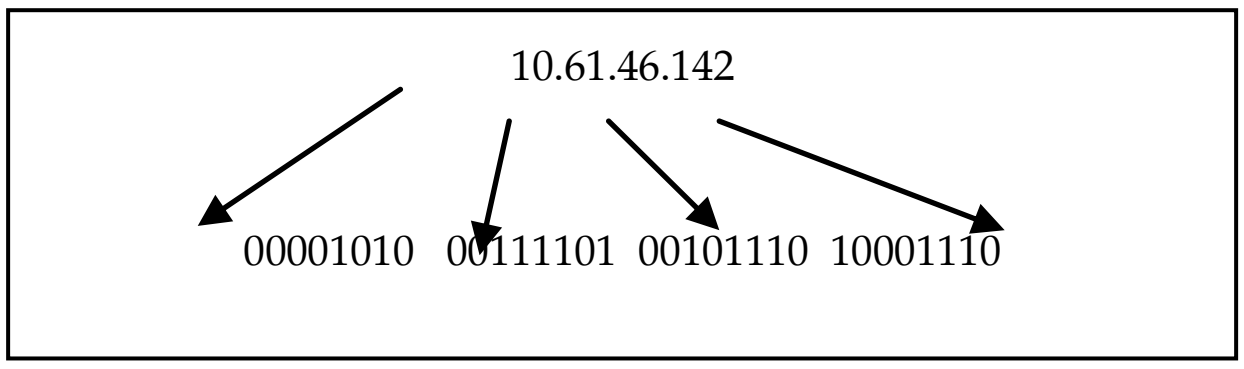

Figure 23 Dotted Decimal Notation and Binary Notation

To run the system, a university campus network is used. The IP Address is Class A. The IP Address available is $2^{24}$ or 16777216. The network address is 10.0.0.0 while the last address is 10.255.255.255.

The first address can be found by ANDing the IP address with the mask. ANDing is done bit by bit. The result of ANDing in figure 24 for 2 bits is 1 if both are 1s; and 0 otherwise.

```
00001010  00111101  00101110  10001110
11111111  00000000  00000000  00000000
--------------------------------------
00001010  00000000  00000000  00000000
```

Figure 24 First Address

The last address can be found by ORing the given addresses with the complement of the mask. ORing here is done bit by bit. The result of ORing in figure 25 for 2 bits is 0 if both bits are 0s; the result is 1 otherwise. The compliment of number is found by changing each 1 to 0 and each 0 to 1.

```
00001010  00111101  00101110  10001110
00000000  11111111  11111111  11111111
--------------------------------------
00001010  11111111  11111111  11111111
```

Figure 25: Last Address

## Motion Detection

For this system, motion based detection technique is used. There are two important elements in the motion detection system; motion sensitivity and motion threshold. The Figure 26 shows the image sequence when a motion is detected.

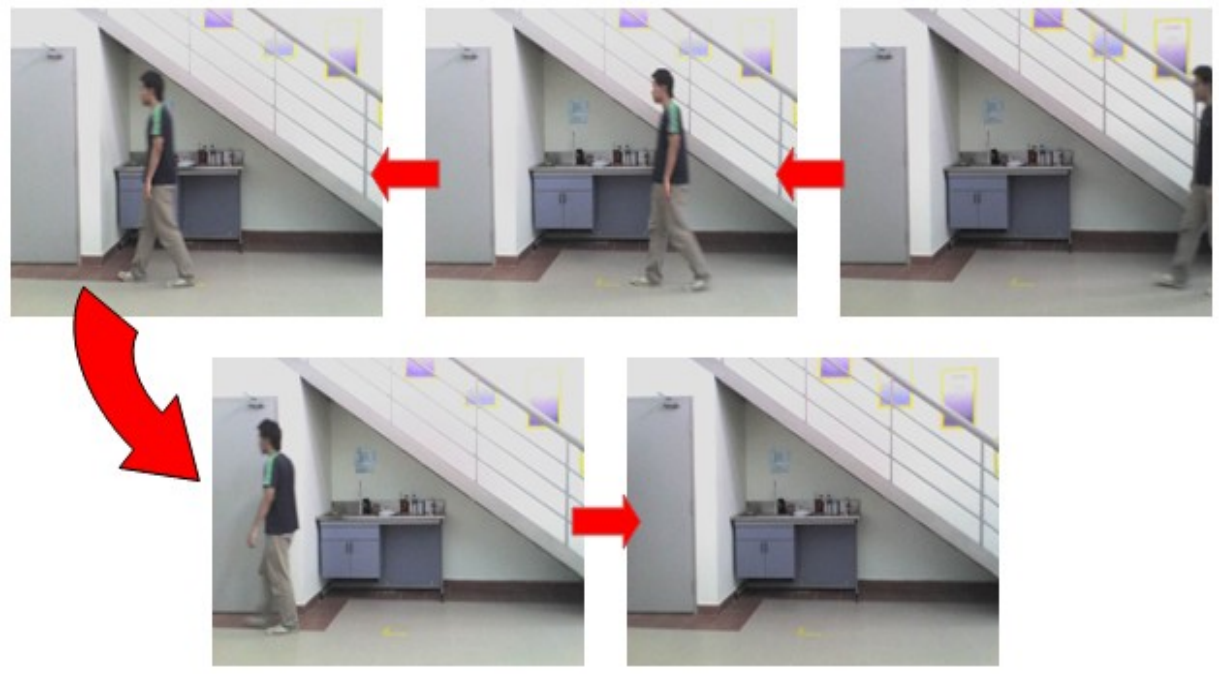

Figure 26 Sequence of Detected Image

## System Level Testing

After each module was completely tested and debugged we integrated the modules to come up with the complete system. Then the entire system is tested

(A) Low threshold

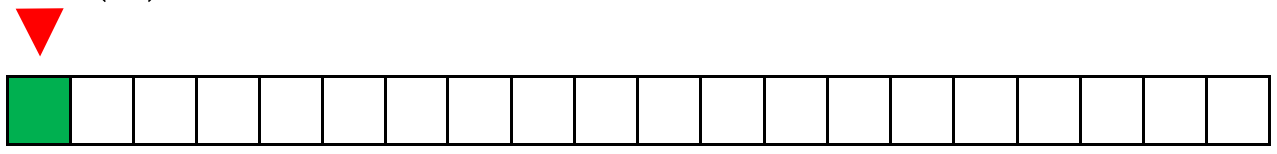

TABLE 3
IMAGE CAPTURED WHEN LOW THRESHOLD SETTING

| Sensitivity / Motion | Low | Moderate | High |
|---|---|---|---|
| **Walk** | 4 image per second | 4 image per second | 5 image per second |
| **Walk Fast** | 3 image per second | 3 image per second | 3 image per second |
| **Run** | No motion is triggered | 2 image per second | 2 image per second |

(B) Moderate threshold

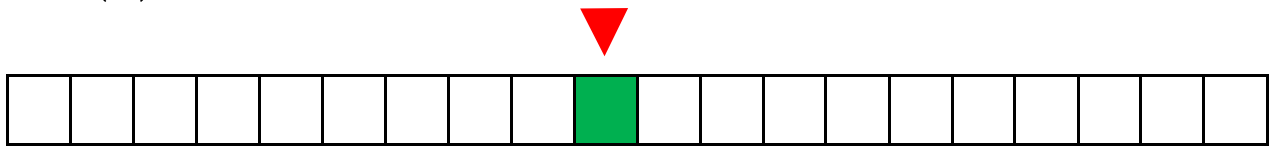

TABLE 4
IMAGE CAPTURED WHEN ODERATE THRESHOLD SETTING

| Sensitivity / Motion | Low | Moderate | High |
|---|---|---|---|
| **Walk** | No motion is triggered | 1 image per second ( distance 3m) | 3 image per second |
| **Walk Fast** | No motion is triggered | No motion is triggered | 2 image per second |
| **Run** | No motion is triggered | No motion is triggered | 2 image per second |

(C) High threshold

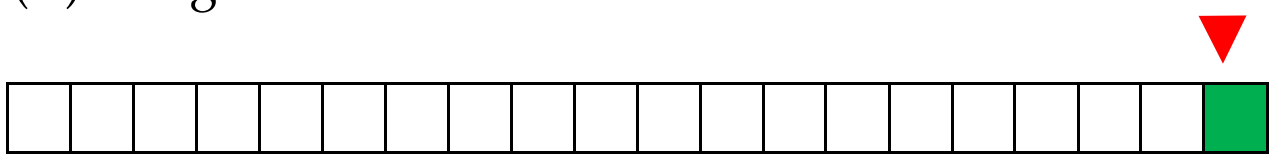

TABLE 5
IMAGE CAPTURED WHEN HIGH TRESHOLD ISSETTING

| Sensitivity / Motion | Low | Moderate | High |
|---|---|---|---|
| **Walk** | No motion is triggered | No motion is triggered | 1 image per second ( distance 1m) |
| **Walk Fast** | No motion is triggered | No motion is triggered | No motion is triggered |
| **Run** | No motion is triggered | No motion is triggered | 2 image per second |

Motion is detected by changes in the outline of objects and changes in object brightness; however, in some cases the camera may detect rapid brightness changes by artificial light sources (such as fluorescent lights) as motion. Motion may not be detected as desired if the object's color is similar to the color of the background. The motion detection feature is disabled when panning and tilting the camera lens, i.e., moving the camera lens will not trigger the motion detection feature. Motion detection can vary by the object, image resolution, or image quality.

Figure 27 shows that the camera can still detect the movement even it only image reflected to the mirror if the sensitivity is setting high.

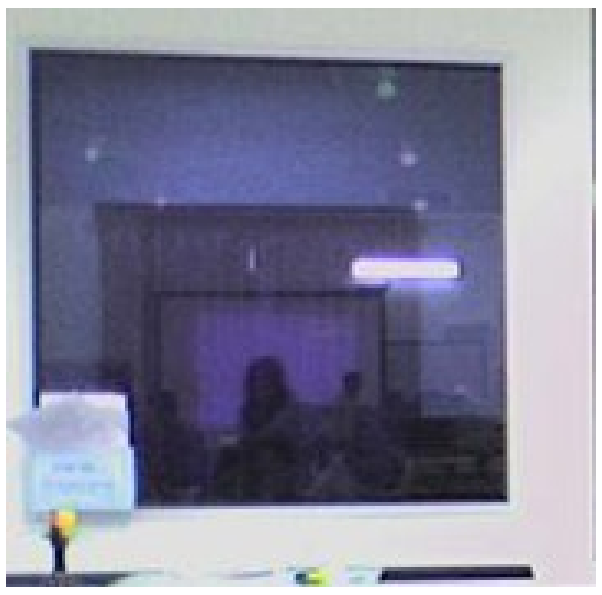

Figure 27 The Image Reflected by the Mirror

Furthermore, when the door is open, the light from outside reflect the camera to detect a motion due to temperature different. Figure 28 shows the image captured when there is a light changes.

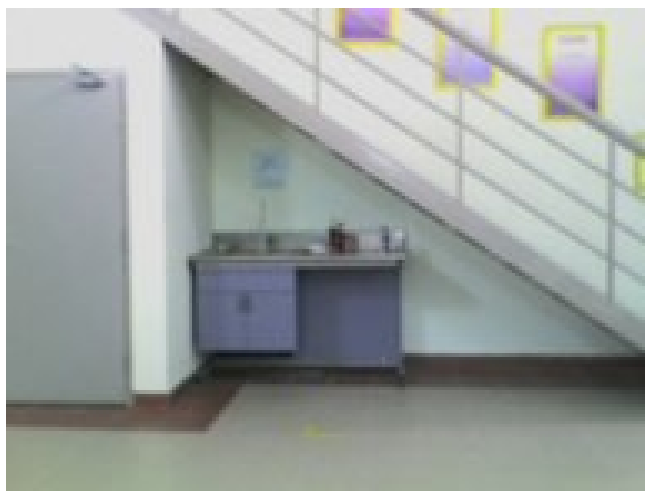

Figure 28 Image of the Light Changes

Under Sensor deactivation time setting, the amount of time that has been set must be passed over prior to detection to be triggered. It is recommended to set the long time interval to avoid image to be buffered. These scenario lead to buffered image is continuously send to mobile phone. Images will not be buffered or transferred during the deactivation time. As a result, pre-buffer images from the next detection may not be buffered or transferred. For example in Figure 29, if this parameter is set to [10s], and if the camera is configured to buffer 1 image per second, the camera will not buffer or transfer images if it is triggered within 10 seconds of the previous trigger. Images can only be buffered or transferred 10 seconds after the previous trigger.

— The camera can be triggered

···· The camera is buffering of transferring images and cannot be triggered during this time.

▪▪▪ Deactivation time, the camera cannot be triggered

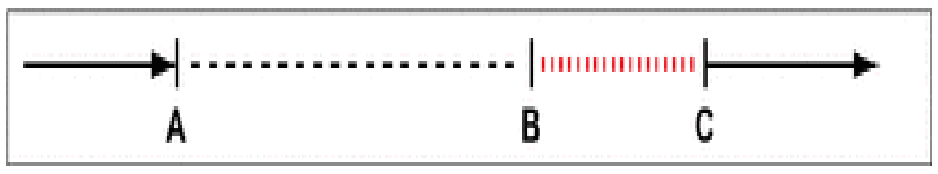

Figure 29: Condition for Image Triggered

For condition A, the camera is triggered. Buffering or transferring begins. No new images will be buffered or transferred. While in condition B, buffering or transferring ends, deactivation time will start and no new images will be buffered or transferred. And for condition C, deactivations time ends, the camera can buffer or transfer images again.

## 4 CONCLUSIONS AND RECOMMENDATIONS

The level of sensitivity and threshold play an important role in motion detection and is required to set an appropriate value before it can detect the motion. The user is automatically notified only when significant changes made to the images such as the appearance of a huge moving object like a human intruder. For the ideal case, the user can set low for threshold and moderate for sensitivity so the best condition can be captured when there is a movement occur.

Besides, a segmentation based on change detection whereby will compare the current image with the reference image (normal state) by comparing images received from the currently selected video capture device with the previous one. Motion detection involves computing the differences between the two images. If the difference between the compared images crosses a certain threshold value then motion is detected.

The system configuration discussed here is simple, feasible and consists of easily accessible and inexpensive components. The evolution of communication technology allows this system developed with easy and user friendly notification system such as SMS.

There are sufficient regular users and awareness behind all services. SMS has become an integral and important part of many customer's everyday business and personal lives. The practicality of this application will results in various benefits to the users from all nature.

The improvement that can be made:

1. Identify the intruder whether human or animal (cat or dog). Normally the traditional way to detect the motion always created false alarm. So combination of image recognition, motion detection, image processing and fuzzy interference can be considered for new method for this system
2. Used HTTP protocol as a transfer method to transfer the image capture by camera. So time of notification received can be shorted because there is no use software such as WinFTP Server (using FTP protocol) to transfer the image.
3. Another method of notification can be used; example e-mail and MMS. Then a comparison can be made between these methods to determine the fastest way.
4. Different type of internet connection can be used such as Streamyx and 3G.
5. The stand-alone program is a Visual Basic executable program, which provides full control of the motion detection system from the actual location. The Visual basic interface will runs in the client computer. The program does not need Internet connection; it has all the functions needed to perform any operation locally

## 5 ACKNOWLEDGMENT

The authors gratefully acknowledge the staffs in the laboratory of Intelligent System and Information Technology Center, University Tun Hussein Onn Malaysia as well as Panasonic, Malaysia for their valuable contributions towards the success throughout this project.